\documentclass[paper ]{revtex4}

\usepackage{graphicx}
\usepackage{rotating}
\usepackage{amsmath}
\usepackage{url}
\usepackage{bm}
\newcommand{\cac}{CaC$_6$}

\newcommand{\pig}{$\pi$}

\newcommand{\be}{\begin{equation}}
\newcommand{\ee}{\end{equation}}
\newcommand{\bea}{\begin{eqnarray}}
\newcommand{\eea}{\end{eqnarray}}
\newcommand{\ba}{\begin{array}}
\newcommand{\ea}{\end{array}}
\newcommand{\Ave}[1] { \langle #1 \rangle }

\newcommand{\EF}{$E_F$}

\newcommand{\nk}{{n\bf k}}
\newcommand{\nkp}{{n'\bf k'}}

\newcommand{\br}{{\bf r}}
\newcommand{\bR}{{\bf R}}

\newcommand{\mgbtwo}{MgB$_2$}
\newcommand{\tc}{$T_{\rm c}$}

\newcommand{\up}{\uparrow}
\newcommand{\down}{\downarrow}

\newcommand{\I}{{\rm i}}
\newcommand{\E}{{\rm e}}

\newcommand{\Tr}[1]{{\rm Tr}\{ #1 \}}
\newcommand{\op}[1]{{\hat #1}}

\newcommand{\mint}[1]{\int\! {\rm d}^{3} #1 \, }
\newcommand{\mdint}[2]{\mint{#1}\!\!\!\mint{#2}}
\newcommand{\KS}{{\rm s}}

\begin{document}


\title{Role of Coulomb interaction in the superconducting properties of  CaC$_6$
and H under pressure}


\author{S. Massidda$^1$, F. Bernardini$^1$, C. Bersier$^3$, A. Continenza$^2$, P. Cudazzo$^2$,  A. Floris$^3$,  H. Glawe$^3$, M. Monni$^1$, S. Pittalis$^3$, G. Profeta$^2$,  A. Sanna$^1$, S. Sharma$^3$,  E.K.U. Gross$^3$}

\affiliation{$^1$SLACS-INFM/CNR,  and Dipartimento di Scienze Fisiche, Universit\`a degli Studi di Cagliari, I-09042
Monserrato (CA), Italy}
\affiliation{$^2$ CNISM - Dipartimento di Fisica,
Universit\`a degli Studi dell'Aquila, Via Vetoio 10,
I-67010 Coppito (L'Aquila) Italy}
\affiliation{$^3$Institut f{\"u}r Theoretische Physik, Freie Universit{\"a}t
Berlin, Arnimallee 14, D-14195 Berlin, Germany}

\begin{abstract}
Superconductivity in intercalated graphite CaC$_6$ and H under extreme pressure, in the framework of superconducting density functional theory, is discussed. 
A detailed analysis on how the electron-phonon and electron-electron interactions combine together to determine the superconducting gap and critical temperature
of these systems is presented. In particular, we discuss the effect on the calculated $T_c$ of the anisotropy of the electron-phonon interaction and of the different approximations for screening the Coulomb repulsion.
These results contribute to the understanding of multigap and anisotropic 
superconductivity, which has received a lot of attention since the discovery 
of MgB$_2$, and show how it is possible to
describe   the superconducting properties of real materials on a fully \emph{ab-initio}
basis. 
\end{abstract}
\maketitle

\section{Introduction}

The discovery of two-gap superconductivity in \mgbtwo\ has triggered a huge revival of 
interest on the search of new electron-phonon superconductors. 
The recent experience in the vast majority of electron-phonon superconductors
has shown that  $ab-initio$  calculations can provide a detailed and useful 
description of the normal state of these systems, including dynamical properties and electron-phonon coupling. The treatment of the superconducting state, however, is 
a much more difficult task. 
On the theoretical side, most of nowadays calculations are based on
the Eliashberg theory \cite{Eliashberg,McMillan,SchScaWil}.
While in the latter the electron-phonon interaction is perfectly accounted for, the
effects of
the electron-electron Coulomb repulsion are condensed in a single parameter, $\mu^*$,
which is difficult to
calculate from first principles and which, in most practical applications,
is treated
as an adjustable parameter usually fitted to the
experimental \tc. In
this sense the Eliashberg theory, in spite of its tremendous success, has to be
considered as a semi-phenomenological theory. However, the possibility to describe the 
superconducting state on a fully $ab-initio$ ground and, as a consequence, to be able to predict the superconducting
properties of materials, is still highly sought for.

The density functional theory (DFT),
a very successful standard approach in normal state electronic structure
calculations, has been recently extended to
deal with the superconducting state (SCDFT) \cite{SCDFT-I,SCDFT-II}. SCDFT contains no adjustable parameters, and the final critical
temperature
is the result of material specific quantities, all of them computed $ab-initio$. SCDFT is able to treat superconductors
with a wide range of couplings, as shown by several investigations, \mgbtwo\  being one of the most non-trivial one \cite{noimgb2}. 

Superconductivity is the
result of a subtle competition between two opposite effects: the phonon
mediated
attraction (``e-ph'' in the following) and the direct Coulomb
repulsion (``e-e'')
between the electrons. In this work, we show how different approximations
for the e-e and e-ph interactions (e.g. an isotropic approximation, or an
insufficiently
elaborated screening entering the e-e term) will depart from the experimental
results. Considering  intercalated graphite CaC$_6$ and H under pressure as
test cases, we will quantify, without ad-hoc assumptions, 
the effect on \tc\ of the superconducting
gap anisotropy and of the assumed screened Coulomb potential.
The choice of these two systems comes from their peculiarities: \cac\ is a superconductor with a moderate degree of
anisotropy \cite{CaC6}, recently confirmed by experiments \cite{cac6_pcs, KimCaC6},
with electronic bands having different orbital characters. 
With a critical temperature as high as  11.5 K, 
\cac\  exhibits the highest \tc\ \cite{Emery} among graphite intercalated compounds. DFT calculations of \cac\ electronic and dynamical 
properties  \cite{CalMauri1,Kim} pointed out 
that the e-ph coupling is large enough to yield the observed \tc. In \cac, the phonons mostly 
contributing to superconductivity belong  to an optical branch involving 
Ca displacements and, to a lower extent, to two C-related branches at much higher frequency. The different phonon branches couple to different Fermi surface (FS) sheets, as pointed out in Ref. \onlinecite{CaC6}.

 High temperature superconductivity
in H was suggested 40 years ago by Ashcroft\cite{ashcroft} and has been the subject of several investigations (see Ref. \onlinecite{noiH} and references therein). Unlike \cac, molecular hydrogen under
pressure involves basically only one type of atomic orbital, it
has very large phonon frequencies (due to the light H mass) and large e-ph coupling \cite{noiH}.
   
The paper is organized as follows: In Section \ref{scdft} we summarize the main features of SCDFT and describe our computational approach; in section \ref{CME} we discuss the different approximations used for the Coulomb interaction; 
in Sections \ref{results_cac6} and  \ref{results_H} we present our results for CaC$_6$ and H, respectively; finally,  in Section \ref{concl},
we summarize our conclusions.

 \section{The Density functional theory for superconductors}
\label{scdft}


The density functional theory  \cite{DreizlerGross} has enjoyed
increasing popularity as a reliable and relatively inexpensive tool to 
describe real materials. In this section we will briefly outline 
the DFT approach to superconductivity, and refer to the original papers for more details. In order to give an introduction to SCDFT, it is  instructive to
recall how magnetism is treated within the DFT.  The Hohenberg-Kohn
(HK) theorem~\cite{HohenbergKohn} states that
all observables, in particular also the magnetization, are functionals of the
electronic density {\it alone}. This, however, assumes the knowledge of the
magnetization as a functional of the density. Finding an
approximation for  this functional is extremely hard and, in practice, one chooses a
different approach. The task can be vastly simplified by treating the
magnetization density ${\bf m}(\br)$, i.e., the order parameter of the magnetic state, as an
additional fundamental density in the density functional framework \cite{hedinvb}.  An
auxiliary field -- here a magnetic field ${\bf B}_{\rm ext}(\br)$ -- is introduced, 
which couples to ${\bf m}(\br)$ and breaks the corresponding (rotational) symmetry 
of the Hamiltonian. This field drives the
system into the ordered state. 
If the system is actually magnetic, the order parameter will survive when the auxiliary perturbation is quenched.
In this way, the ground-state magnetization density is determined by minimizing
the total energy functional (free energy functional for finite temperature
calculations) with respect to both the normal density and the magnetization
density.  Within this  approach a much simpler approximations to the xc functional (now a functional of
two densities) can lead to satisfactory results.

The same idea is also at the heart of density functional theory for
superconductors, as formulated by Oliveira, Gross and Kohn~\cite{ogk}.  Here the
order parameter is the so-called anomalous density,
\begin{equation}
  \label{eq:def-chi}
  \chi(\br,\br') = \Ave{\hat{\Psi}_{\up}(\br) \hat{\Psi}_{\down}(\br')} \, ,
\end{equation}
and the corresponding potential is the non-local pairing potential
$\Delta(\br,\br')$. It can be interpreted as an external pairing field, induced
by an adjacent superconductor via the proximity effect. Again, this external
field only acts to break the symmetry (here the gauge symmetry) of the system,
and is quenched at the end of the calculation. As in the case of magnetism,
if the system is actually a superconductor the order parameter will be sustained by the self-consistent 
effective pairing  field.
The approach outlined so far captures, in principle, all the electronic degrees of
freedom.  To describe conventional phonon-mediated superconductors, also the
electron-phonon interaction has to be taken into account.

In  order to treat both weak and strong
electron-phonon coupling, the electronic and the nuclear degrees of freedom
have to be treated on equal footing. This can be achieved by a multi-component
DFT, based on both the electronic density and the nuclear
density~\cite{kreibich,kreibich2008}.  
 In addition to the normal and anomalous
electronic densities, we also include the diagonal of the nuclear density matrix
\footnote{Taking only the nuclear density would lead to a system of strictly
  non-interacting nuclei which obviously would give rise to non-dispersive,
  hence unrealistic, phonons.}

\begin{equation}
  \label{eq:def-Gamma}
  \Gamma(\underline{\bR}) = 
  \Ave{\hat\Phi^\dagger(\bR_1) \dots \hat\Phi^\dagger(\bR_N)
    \hat\Phi(\bR_N) \dots \hat\Phi(\bR_1)},
\end{equation}
where $\hat{\Phi}(\bR)$ is a
nuclear field operator.

In order to formulate a Hohenberg-Kohn theorem for this system, we introduce a
set of three potentials, which couple to the three densities described above. Since
the electron-nuclear interaction, which in conventional DFT constitutes the external
potential, is treated explicitly in this formalism, it is {\em not} part of the
external potential. The nuclear Coulomb interaction $\hat{U}^{\rm nn}$ already has the 
form of an external many-body potential, coupling to $\Gamma(\underline{\bR})$, and
for the sake of the Hohenberg-Kohn theorem, this potential will be allowed to take
the form of an arbitrary N-body potential.
All three external potentials are merely mathematical
devices, required to formulate a Hohenberg-Kohn theorem. At the end of the
derivation, the external electronic and pairing potentials will be set to zero while 
the external nuclear many-body potential to the nuclear Coulomb interaction.

As usual, the Hohenberg-Kohn theorem guarantees a one-to-one mapping between the
set of the densities $\{n(\br),\chi(\br,\br'),\Gamma(\underline{\bR})\}$ in
thermal equilibrium and the set of their conjugate potentials $\{v_{\rm
  ext}^\text{e}(\br)-\mu,\Delta_{\rm ext}(\br,\br'),v_{\rm
  ext}^\text{n}(\underline{\bR})\}$.  Therefore all the observables are
functionals of the set of densities. Finally, it assures that the grand canonical
potential,
  \be
  \label{eq:intomega}
  \Omega[n,\chi,\Gamma] = F[n,\chi,\Gamma] + \mint{r} n(\br) 
  [v^\text{e}_{\rm ext}(\br) - \mu]    
  - \mdint{r}{r'} \left[ \chi(\br,\br') \Delta_{\rm ext}^*(\br,\br') + \text{h.c.} \right]  
  + \mint{\underline R} \, \Gamma(\underline{\bR}) v_{\rm ext}^\text{n}(\underline{\bR})
  ,
  \ee
is minimized by the equilibrium densities. We use the notation $A[f]$ to denote that
$A$ is a functional of $f$. The functional $F[n,\chi,\Gamma]$ is
universal, in the sense that it does not depend on the external potentials, and
is defined by
\be
  \label{eq:intF}
  F[n,\chi,\Gamma] = T^\text{e}[n,\chi,\Gamma] + T^\text{n}[n,\chi,\Gamma]   
  + U^\text{en}[n,\chi,\Gamma] + U^\text{ee}[n,\chi,\Gamma]
  - \frac{1}{\beta} S[n,\chi,\Gamma]
  \,,
\ee
where $S$ is the entropy of the system,
\begin{equation}
  S[n,\chi,\Gamma] = -\Tr{\op\rho_0[n,\chi,\Gamma] \ln(\op\rho_0[n,\chi,\Gamma])}
  \,.
\end{equation}

In standard DFT one normally defines a Kohn-Sham system, i.e., a non-interacting
system chosen such that it has the same ground-state density as the interacting
one. The variational procedure for this system gives   Schr{\"o}dinger-like (Kohn-Sham) 
equations for non-interacting electrons subject to an effective (Kohn-Sham) potential.
These equations are nowadays routinely solved by solid state theorists. 
In our  formalism, the Kohn-Sham system consists of non-interacting
(superconducting) electrons, and {\it interacting} nuclei. We will not  describe here the 
details of the method, and will only outline its basic features:
The Kohn-Sham potentials, which are derived in analogy to normal DFT, include
the external fields, Hartree, and exchange-correlation terms. The latter account
for all many-body effects of the electron-electron and electron-nuclear
interactions. Obtaining  their explicit form has represented a major theoretical 
effort\cite{KurthPhd,LuedersPhd,MarquesPhd}. 
Once this problem has been solved, 
the problem of minimizing the Kohn-Sham grand canonical
potential
 can be transformed into a set of three
differential equations that have to be solved self-consistently: One equation
for the nuclei, which resembles the familiar nuclear Born-Oppenheimer equation,
and two coupled equations which describe the electronic degrees of freedom and
have the algebraic structure of the Bogoliubov-de Gennes~\cite{Bogoliubov58}
equations.

The resulting Kohn-Sham Bogoliubov-de Gennes (KS-BdG) equations read (we use atomic Rydberg units)

\begin{subequations}
  \label{KS-BdG}
  \begin{align}
    \left[ -\frac{\nabla^2}{2} + v^\text{e}_\KS(\br) - \mu \right]
    u_\nk(\br)   
    +\mint{r'} \Delta_\KS(\br,\br') v_\nk(\br') &= \tilde{E}_\nk \, u_\nk(\br) \,,   \\
    - \left[ -\frac{\nabla^2}{2} + v^\text{e}_\KS(\br) - \mu \right]
    v_\nk(\br) 
    + \mint{r'} \Delta^*_\KS(\br,\br') u_\nk(\br') &= \tilde{E}_\nk \, v_\nk(\br)\,, 
  \end{align}
\end{subequations}
where $u_\nk(\br)$ and $v_\nk(\br)$ are the particle and hole amplitudes.  This
equation is very similar to the Kohn-Sham equations in the OGK
formalism~\cite{ogk}.  However, in the present formulation the lattice potential
is not considered an external potential but enters via the electron-ion
Hartree term. Furthermore, our exchange-correlation potentials depend on the
nuclear density matrix, and therefore on the phonons. Although Eq.
 (\ref{KS-BdG}) and 
 the corresponding equation for the nuclei
have the structure of static mean-field equations, they contain, in principle,
all correlation and retardation effects through the exchange-correlation
potentials.  

These KS-BdG equations can be simplified by the so-called decoupling
approximation~\cite{GrossKurth91,SCDFT-I}, which corresponds to the following
ansatz for the particle and hole amplitudes:
\begin{equation}
  \label{eq:decapprox}
  u_\nk(\br) \approx u_\nk \varphi_\nk(\br)
  \,; \quad
  v_\nk(\br) \approx v_\nk \varphi_\nk(\br)
  \,,
\end{equation}
where the wave functions $\varphi_\nk(\br)$ are the solutions of the normal
Schr{\"o}dinger equation.  In this way the eigenvalues in Eqs.~(\ref{KS-BdG})
become $\tilde E_\nk = \pm E_\nk$, where
\begin{equation}
  E_\nk = \sqrt{\xi_\nk^2+|\Delta_\nk|^2} \,,
\end{equation}
and $\xi_\nk = \epsilon_\nk-\mu$. This form of the eigenenergies allows us to
interpret the pair potential $\Delta_{\nk}$ as the gap
function of the superconductor. Furthermore, the coefficients $u_\nk$ and $v_\nk$ 
are given by simple expressions within this approximation
\begin{subequations}
  \begin{align}
    u_\nk & = \frac{1}{\sqrt{2}}{\rm sgn}(\tilde E_\nk) \E^{\I\phi_\nk}
    \sqrt{1+\frac{\xi_\nk}{\tilde E_\nk}} \,,
    \\
    v_\nk & = \frac{1}{\sqrt{2}} \sqrt{1-\frac{\xi_\nk}{\tilde E_\nk}}
    \,.
  \end{align}
\end{subequations}
Finally, the matrix elements $\Delta_\nk$ are defined as
\begin{equation}
  \label{eq:delta_i}
  \Delta_\nk = \mdint{r}{r'} \varphi^*_\nk(\br)\Delta_\KS(\br,\br')\varphi_\nk(\br')
  \,,
\end{equation}
and $\phi_\nk$ is the phase $\E^{\I\phi_\nk} = \Delta_\nk/|\Delta_\nk|$.  The
normal and the anomalous densities can then be easily obtained from:
\begin{subequations}
  \label{eq:el-densities}
  \begin{align}
    n(\br) & = \sum_\nk \left[1-\frac{\xi_\nk}{E_\nk}
      \tanh\left(\frac{\beta}{2}E_\nk \right)\right]
    |\varphi_\nk(\br)|^2
    \\
    \chi(\br,\br') & = \frac{1}{2}\sum_\nk
    \frac{\Delta_\nk}{E_\nk} \tanh\left(\frac{\beta}{2}E_\nk \right)
    \varphi_\nk(\br)\varphi^*_\nk(\br')
    \,.
  \end{align}
\end{subequations}
Within the decoupling approximation, we finally arrive at an equation for the 
 {\bf k}-resolved superconducting gap $\Delta_{n{\bf k}}$, which  has the following form\cite{ogk,SCDFT-I,SCDFT-II,PsikRev}:
\begin{equation}
  \label{eq:gap} 
  \Delta_\nk = - {\cal Z}_\nk \Delta_\nk -\frac{1}{2}\sum_\nkp
  {\cal K}_{\nk,\nkp} \frac{\tanh\left(\frac{\beta}{2}E_\nkp\right)}{E_\nkp} 
  \Delta_\nkp
  \,.
\end{equation}

Eq. (\ref{eq:gap}) is the central equation of the DFT for superconductors. The kernel
 ${\cal K}$  consists of two contributions 
${\cal K}={\cal K}^{\rm e-ph}+{\cal K}^{\rm e-e}$, representing the effects of 
the e-ph and of the e-e interactions, respectively. The diagonal term ${\cal Z}$ plays a similar role as the renormalization term in the Eliashberg 
equations. Explicit expressions  
of ${\cal K}^{\rm e-ph}$ and ${\cal Z}$, which are the results of
the approximate functionals, are given in Eqs. 9 and 11 of Ref.~\cite{SCDFT-II} 
respectively. These two terms
involve the e-ph coupling matrix, while ${\cal K}^{\rm e-e}$ contains the matrix elements 
of the screened Coulomb interaction (the explicit expression is given below). Eq.~(\ref{eq:gap}) has the same structure 
as the BCS gap equation, with the kernel ${\cal K}$ replacing the model interaction 
of BCS theory. This similarity allows us to interpret the kernel as an effective 
interaction responsible for the binding of the Cooper pairs.  Moreover, we emphasize that Eq.~(\ref{eq:gap}) is not a mean-field equation (as  
in BCS theory),  since it contains correlation effects via the SC exchange-correlation functional entering ${\cal K}$ and ${\cal Z}$. Furthermore, it has the 
form of a static equation -- i.e., it does not depend {\it explicitly} on the frequency --  
and therefore has a simpler structure (and computationally more manageable)  than the Eliashberg equations. However, this 
certainly does not imply that retardation effects are absent from the theory.  Once again, retardation effects
enter through the xc functional, as explained in Refs. ~\cite{SCDFT-I,SCDFT-II} . The SCDFT allows to treat on the same footing the e-ph and the screened e-e interactions.
These terms, however, can be treated at different  levels of approximation.  

 We calculated the screened Coulomb matrix elements (ME) with respect to the Bloch functions, for the whole energy range 
of relevant valence and conduction states.  The different 
nature of the  electronic  bands in each material ($e.g.$  some of them can be highly localized 
while others more delocalized), strongly calls for the use of a non-diagonal screening, 
including local field effects.  In order to properly describe these effects, a very important step
 to achieve good agreement with the experiment, we calculated the static random phase approximation (RPA)  
dielectric matrix (DM) $\epsilon^{-1}\!\left({\bm q},{\bm G},{\bm G}'\right)$, using
the pseudopotential-based SELF code \cite{SELF}. The explicit expression of the  
kernel ${\cal K}^{\rm e-e}$ in reciprocal space reads:
\be
\label{eq:rpame}
{\cal K}_{n {\bm k},n' {\bm k'}}^{\rm e-e}=4\pi \sum_{{\bm G},{\bm G}'} \epsilon^{-1}\!\left({\bm q},{\bm G},{\bm G}'\right) 
\frac{\left< n'{\bm k}'|e^{i\left({\bm q}+{\bm G}\right)\cdot {\bm r}}|n {\bm k} \right >\left < n{\bm k}|e^{-i\left({\bm q}+{\bm G}'\right)\cdot {\bm r}}|n'{\bm k}'\right>}{\left|{\bm q}+{\bm G}\right|\left|{\bm q}+{\bm G}'\right|},
\ee
where ${\bm q}={\bm k'}- {\bm k}$. The detailed structure, energy and {\bf k}-dependence 
of the kernel ${\cal K}^{\rm e-e}$ changes the Coulomb renormalization effect 
(due to the different scales of the vibrational and electronic energies) considered: 
In many materials, it only acts as a scaling of the superconducting gap at the Fermi energy, thus
keeping the main structures of the coupling (See Ref. \cite{CaC6,Pb,lial,K}). On the other hand, in strongly anisotropic materials with small interband interaction (like \mgbtwo), the Coulomb renormalization turns out to be 
non trivial\cite{noimgb2,MgB2-Review}.


In this paper we report a study of the role of the different approximations for the Coulomb and the electron-phonon  interactions on
the solution of the SCDFT gap equation. This is illustrated in \cac\cite{CaC6} and H under extreme pressure.
As mentioned above, the  normal state calculations,  necessary for the study  of the 
superconducting state,  are performed within DFT in the LDA or GGA approximations.
Computationally, the electronic and  dynamical properties are obtained using the
pseudopotential method as implemented in the QUANTUM-ESPRESSO package\cite{pwscf};
 the screened Coulomb matrix elements are obtained with the SELF\cite{SELF} code.

\section{Evaluation of Coulomb Matrix elements}
\label{CME}
In this section we discuss the different approximate formulations of the Coulomb interaction analyzed in this work. 

\subsection{Sham-Kohn}
The first and simplest approximation includes a Thomas-Fermi (TF) dielectric function together with free-electron wavefunctions. This leads to isotropic (i.e. {\bf k} independent) matrix elements, expressed by the analytic formula\cite{SCDFT-I,SCDFT-II}:
\begin{equation}\label{eq:ME-TF-KS}
K^{e-e}_{\xi,\xi'}=\frac{\pi}{2\sqrt{\xi \xi'}}{\rm ln}\left[\frac{\xi+\xi'+2\sqrt{\xi\xi'}+ q^2_{TF}/2}{\xi+\xi'-2\sqrt{\xi\xi'}+ q^2_{TF}/2}\right],
\end{equation}
where the Fermi level is fixed by the number of valence electrons and the TF wavevector is that of a jellium having the same density of states at the Fermi energy of the true system $N(0)$, i.e. $q^2_{TF}=8\pi N(0)$ . A justification for the use of this approximation, in the spirit of the seminal paper of Sham and Kohn \cite{Sham-Kohn-V}, is given in Ref.~\cite{SCDFT-I}. Typically, Eq. (\ref{eq:ME-TF-KS}) provides reasonable results for systems with delocalized electrons \cite{SCDFT-II,lial} . Furthermore, it avoids 
the cumbersome calculation of the anisotropic Coulomb matrix elements, calculated with the Kohn-Sham Bloch functions of the system (see below).
 
\subsection{Bloch wavefunctions and Thomas-Fermi} 

In order to include the effects of the wavefunction localization in the evaluation of the matrix elements, we used the real 
Kohn-Sham Bloch states of Eq. (\ref{eq:rpame}), together with the TF screened dielectric function. The TF vector is again  $q^2_{TF}=8\pi N(0)$, 
where $N(0)$ is the density of states at the Fermi energy in the real material.
In this way, we correct for the locality of the wavefunctions in the evaluation of the Coulomb integrals, still keeping, for the screening, a very approximate form.  The present   approximation and the ones described below in this section correspond to an anisotropic treatment of the Coulomb interaction. 
 
\subsection{Diagonal random phase approximation (RPA)}

The TF dielectric function is only valid in the limit ${\bf q}\to0$. 
To improve this aspect we calculate the dielectric function $\epsilon\!\left({\bm q},{\bm G},{\bm G}'\right)$ in the RPA, initially limiting ourselves to the diagonal form, 
where we set to zero all the $\bf{G} \neq \bf{G'}$ terms.
In real space this corresponds to a screening depending not on ${\bf r}$ and ${\bf r}'$ separately 
but only on ${\bf r}-{\bf r}'$. On the other hand, the electronic polarizability is built by taking into account  the real band structure of the system. 
This approximation is expected to be reasonable for
 closely packed systems  with delocalized electrons.

\subsection{Non-diagonal RPA}

With a fully non local RPA screening (with non-zero $\bf{G} \neq \bf{G'}$ terms) the electronic response will depend on the two spatial indices ${\bf r}$ and ${\bf r}'$ separately. This improves the efficiency of the screening for strongly localized states, with a resulting reduction of the Coulomb matrix elements in the corresponding regions. 
As an example, this improvement affects mainly the graphene sheets in \cac.


\section{Results on $\rm CaC_6$}\label{results_cac6}

\begin{figure}
\begin{center}
\begin{minipage}{0.5\textwidth}
\begin{center}
  \includegraphics[clip= ,width=\textwidth]{figs/bande_spd_1.eps}
\end{center}
\end{minipage}
\begin{minipage}{0.2\textwidth}
\begin{center}
 \includegraphics[clip= ,width=\textwidth]{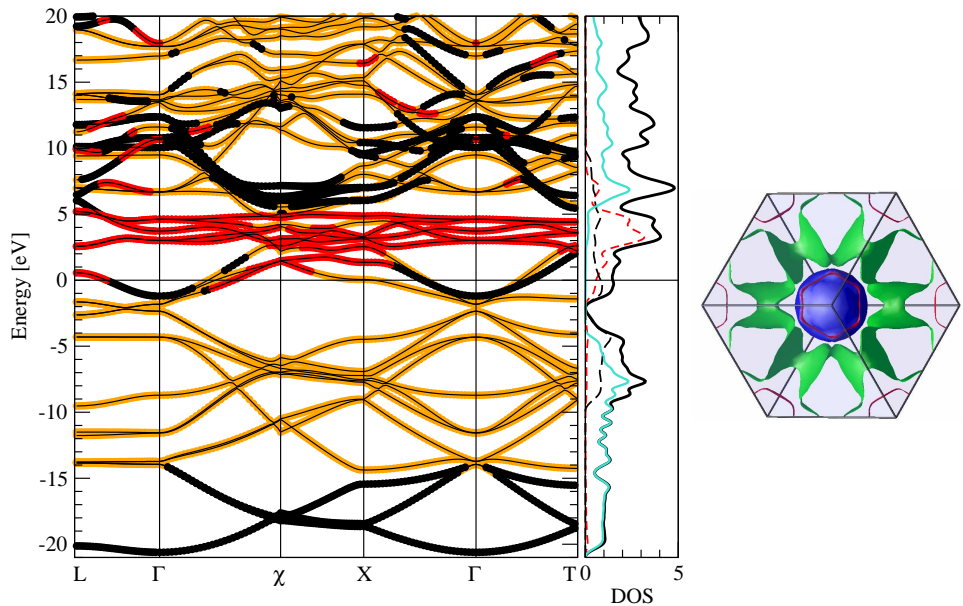}
%
\end{center}
\end{minipage}

\begin{minipage}{0.8\textwidth}
\caption{Left panel: Electronic band structure of \cac, with the Fermi energy set to zero. 
Different color for the Kohn-Sham eigenvalues are used corresponding to the $s$ (black) $p$ (orange-light grey) or $d$ (red-gray) main orbital character. 
Central panel: Total density of states (DOS) (thick black line), projected DOS on Ca site (dashed red-gray line) and on C states, divided in $sp^2$ part (turquoise-light gray), and $p_z$ part (long dashed black).
Right panel: \cac\ Fermi surface, showing the external $\pi$ sheet (1, green), the Ca-related sphere (2, blue) and the internal $\pi$ sheet (3, red).  
}\label{fig:CaC6bands}
\end{minipage}
\end{center}
\end{figure}

Before discussing the various approximations for the Coulomb interaction, we briefly describe  the
electronic states of \cac. This material crystallizes in the rhombohedral space group $\rm R\bar3m$
\cite{Emery}, and has one unit formula per primitive cell. The rhombohedral lattice parameter is 5.17 \AA\
with rhombohedral angle 49.55$\rm ^o$. 
We plot in Fig.~\ref{fig:CaC6bands} the band structure of the system. 
Three bands cross the Fermi energy. Two of them have $p$ character and come from the \pig\ bonding pattern of the graphene layer. They give rise to two  2-dimensional Fermi surfaces (Fig.~\ref{fig:CaC6bands}, right panel).
The third band corresponds to a spherical Fermi surface with the charge localized in the interlayer space and on the Ca ion. Its character is mainly $s$  ($d$) in the $k_x,k_y$ ($k_z$) directions ($z$-axis being normal to the graphene layer).

The presence of an intercalant band crossing the Fermi energy has been shown to be a necessary condition for superconductivity in graphite intercalated compounds (GICs), as pointed out by Cs\'anyi \textit{et al.} \cite{Csanyi}. 
Interlayer bands, uncorrelated  to a particular atomic orbital nature, have been largely discussed in graphite\cite{Baldereschi}, GICs\cite{Csanyi,MazinPRL,CalMauri1,CalMauri2} and MgB$_2$ \cite{noimgb2}.
In CaC$_6$, however, this band gains  a further strong contribution from the Ca $4s$ and $3d$ orbitals. This band is coupled to the in-plane Ca optical phonon branch; these  low frequency modes provide a strong contribution to the e-ph constant $\lambda$ \cite{CalMauri1,Kim,CaC6}.
The energy window between -20 and -4 eV is mainly occupied by the bands forming the $sp^2$ bonds pattern of graphene, as shown by the orbital decomposed density of states in Fig.~\ref{fig:CaC6bands}. The $p_z$ bands starts around the L symmetry point (corresponding to the A point in the hexagonal unit cell, i.e. to the zone boundary in z direction) at about -10 eV.
Just above the Fermi level, mainly between 2 and 5 eV the DOS shows a peak originated by a set of flat bands due to Ca $d$ states. 
Above 5 eV the percentage of the total charge projected over atomic orbitals starts to reduce corresponding to the free electron limit of the KS-wavefunctions.



\subsection{Anisotropic approach for the Coulomb interaction}

We now discuss the effect of the  approximate treatments of the Coulomb interaction described in Sec. \ref{CME} on the superconducting \tc. In the following analysis, we assume a fully anisotropic phononic kernel. 
Within the Sham-Kohn approximation we obtain \tc=13.5 K, to be compared with the experimental
value $T^{\rm exp}_c$=11.5 K. 
The Sham-Kohn approximation is rather simple and gives quite reliable results for nearly free electron metals \cite{SCDFT-II,lial}. 
However, for \cac\ this approximation is not expected to hold, since it neglects the presence of localized states with much stronger Coulomb interactions. Therefore in the Sham-Kohn approximation an overestimate of 
the superconducting gap and $T_c$  is expected and found.

With the Thomas-Fermi approximation and Bloch wavefunctions we obtain $T_c\approx$ 8.2 K, underestimating the experimental value. Remarkably, 
the  critical temperature obtained using the RPA approximation with diagonal screening turns out to be $T_c\approx$ 8.0 K,  very close to the one found in the TF approach and   
quite smaller than the experimental value. 
This result shows that the TF screening is quite similar to the static diagonal RPA one, 
and that indeed the \tc\ underestimation is related to the lack of local field corrections. A diagonal screening
in fact  averages out over a unit cell of the system. A correct  screening, on the other hand, should be
more effective in regions of the unit cell with larger charge density. 
As a consequence,  inclusion of non-diagonal terms in the screening  produces an increased \tc\ of 9.4 K. This value, while 
not in excellent agreement with the experiment, represents an improvement over the previous \tc\ values  (we recall that in our calculations no fitting parameter to the experiment is used).
  In the Table below  we resume the superconducting properties in the discussed cases,
including \tc\ and the value of the gap $\Delta_{nk}$ on the different Fermi surface sheets.

\begin{center}
\begin{minipage}{0.5\textwidth}
$\begin{array}{|c|c|c|c|c|c|}  
\hline
                    & SK-TF  &wf-TF & diag-RPA & RPA   &{\rm no\;Coulomb}\\ \hline
\left<\Delta_1\right>[meV]&   2.20 &1.24  &   1.19   & 1.42  &  5.40 \\  
\left<\Delta_2\right>[meV]&   2.94 &1.78  &   1.77   & 2.00  &  6.72 \\ 
\left<\Delta_3\right>[meV]&   2.49 &1.39  &   1.37   & 1.67  &  6.08 \\ \hline
         T_c [K]          &  13.5  &8.2   &   8.0    & 9.4   &   31  \\ 
\hline
\end{array}$
\end{minipage}
\end{center}

\noindent The brackets $<\cdots>$ indicate a Fermi surface average and the indices $1$,$2$,$3$ correspond to the three parts of the Fermi surface (see Fig.~\ref{fig:CaC6bands}), namely the external (green)
\pig\ surface, the Ca sphere (blue) and the internal (red) \pig\
surface. 

\begin{figure}[htbp]
\begin{center}
\begin{minipage}{0.7\textwidth}
\includegraphics[clip= ,width=\textwidth]{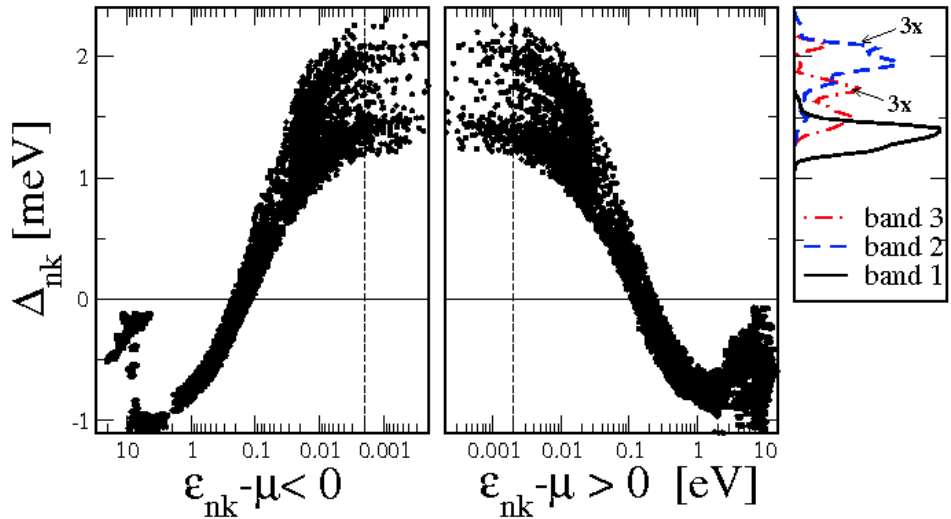}
\caption{\cac\ superconducting gap $\Delta_{n{\bf k}}$ as a function of the energy distance from the Fermi energy ($\mu$ in the figure). The right panel shows the energy distribution (in arbitrary units) of
the  gap, resolved over the three Fermi surfaces. The histogram is
evaluated within the energy window from -2  to 2 meV.}
\label{fig:gapcac6}
\end{minipage}

\end{center}
\end{figure}

The detailed behavior of the gap $\Delta_{n{\bf k}}$ as a function of the normal state eigenvalue is given in Fig.~\ref{fig:gapcac6}. The  set of black points shows the gap at $T=0$~K, as a function of the energy distance from the Fermi level \EF. For each energy the gap is not a single-valued function, i.e. the gap is anisotropic in the reciprocal space. In particular, the gap values exhibit a moderate anisotropy, rather than a multigap character.
 
\subsection{Isotropic approach for the Coulomb interaction}
All the above results (excluding the Sham-Kohn approximation) refer to fully anisotropic calculations, in which the ${\bf k}$, ${\bf k}'$ 
dependence of the Coulomb matrix elements is included in the solution of the SCDFT gap equation. 
We also calculated \tc\ with an isotropic approach in which the ${\cal K}_{n {\bm k},n' {\bm k'}}^{\rm e-e}$  
matrix elements are averaged in ${\bf k}$ and ${\bf k}'$ over isoenergy surfaces, yielding a corresponding $V\left(\xi,\xi'\right)$ function. The average is done for each one of the approximations discussed. 
The critical temperatures turn out to be quite close (of the order of few percent) to the corresponding fully anisotropic ones, 
showing that the anisotropy effect on the Coulomb repulsion is almost negligible,  unlike in the case of \mgbtwo \cite{noimgb2,MgB2-Review}.
On the other hand (see Sec. \ref{effph}), this is not true for the phononic part of the kernel \cite{CaC6,cac6_pcs,KimCaC6}.

\begin{figure}
\begin{center}
\begin{minipage}{0.8\textwidth}
\begin{center}
 \includegraphics[clip= ,width=\textwidth]{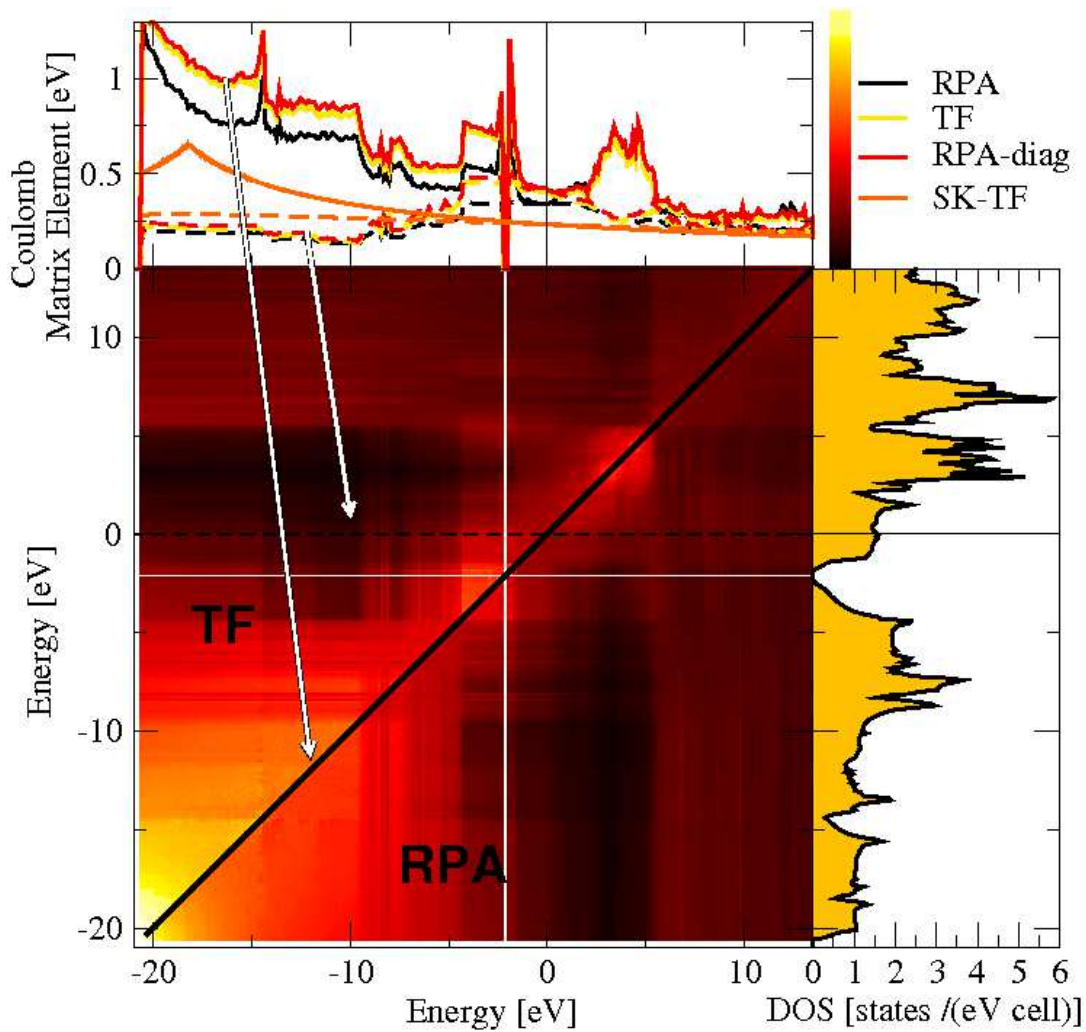}
\end{center}
\end{minipage}

\begin{minipage}{0.8\textwidth}
\caption{Comparison between different approximations in calculating the Coulomb matrix elements (ME). 
The central panel shows, with a scale of colors, the matrix of averaged Coulomb ME (defined in Eq.~\ref{eq:rpame}). Because this matrix is symmetric, we used the upper-left part to show Thomas-Fermi ME (using Bloch wavefunctions) and the bottom-right part for RPA ME.
The upper panel shows two cuts of this matrix for the various approximations discussed in the text: A diagonal cut [$V\left(\xi,\xi\right)$, straight lines] and an horizontal cut passing for E$_F$ [$V\left(0,\xi'\right)$, dashed lines]. On the right the density of electronic states.}\label{fig:mappa}
\end{minipage}
\end{center}
\end{figure}

A graphical comparison between the (averaged) different approximations is given in Fig.~\ref{fig:mappa}, where we report (upper panel) two cuts of the  function $V\left(\xi,\xi'\right)$, namely the diagonal part $V\left(\xi,\xi\right)$ and the elements with the first energy corresponding to the Fermi level $V\left(0,\xi'\right)$; in the lower panel a graphical representation of the whole function in the TF  and in RPA approximation is presented.
The Sham-Kohn approach gives a smooth function, with no correlation with the electronic structure of the material. The upturn at about -18 eV corresponds to the energy of the ${\bf k}=0$ state in a free electron gas with the same number of electrons per unit cell. 
The use of TF approximation with the Bloch 
wavefunctions shows the effect of the orbital nature of states reflected in the matrix elements. In Fig.~\ref{fig:mappa} for example we see the strong Coulomb repulsion between very localized Ca-$d$ states giving a peak in the Coulomb matrix elements for $\xi \simeq \xi' \simeq$ 4 eV. The strong localization implies a nearly zero interaction between these states and the rest of the electronic structure (the darkest regions in the matrix), whose charge is located in different areas . Moreover we can distinguish the signature of the $sp^2$ bonds in the region -20 to -5 eV in a sharp increase of the matrix elements with poor interaction with higher energy states. 
 
Going from Thomas-Fermi to diagonal-RPA  screening does not introduce any important changes. Local fields effects, on the other hand, strongly reduce the matrix elements below the Fermi level. 
In the high energy region all the approximations behave in a similar way as the states are free-electron like. The RPA including local field effects, although giving similar structures with respect to the Bloch wavefunctions and TF approach,  introduces a more efficient screening in the graphene layer so that the corresponding matrix elements are reduced of about 20\%.

\subsection{Effects of anisotropy in the phonon coupling}
\label{effph}

It is interesting, at this point of the discussion, to analyze the role of the e-ph interaction anisotropy and its effects on \tc\ and on the anisotropic character of the superconducting gap.
 
 From a multiband BCS model (i.e. the Suhl-Matthias-Walker model  \cite{SMW}), it is known that an anisotropy in the phonon coupling always yields a \tc\ enhancement with respect to an averaged coupling.
To discuss this kind of anisotropy, it is useful to compare \cac\ with \mgbtwo, the prototypical two gap superconductor. We first define the band resolved e-ph coupling $\lambda_{ij}=\int \alpha^2F_{ij}(\omega)/\omega d\omega$ and Eliashberg functions $\alpha^2F_{ij}$
\begin{equation}\label{eq:definea2Fij}
\alpha^2F_{ij}(\omega) =  \frac{1}{N_i}\sum_{{\bf k},{\bf k}'}\sum_{\nu}|g^{ij}_{{\bf k},{\bf k}',\nu}|^2\delta(\xi_{i{\bf k}})\delta(\xi_{j{\bf k}'})\delta(\omega-\omega_{{\bf q}\nu}),
\end{equation}
where: $i$ and $j$ are indices of the selected sheet of the Fermi surface;  $N_j$ is partial DOS at Fermi level, for the sheet $j$; $|g^{ij}_{{\bf k},{\bf k}',\nu}|^2$ are the e-ph matrix elements; $\nu$ is the phonon branch; $\omega_{{\bf q}\nu}$ the phonon frequencies. Moreover, we define the DOS-renormalized, band resolved matrix elements
$V^{ph}_{ij}=\lambda_{ij}/N_j$.

In the following, we present the partial DOS and couplings for CaC$_6$ and MgB$_2$: For CaC$_6$ we separate the Fermi surface in three parts: \textsl{(1)} external $\pi$ bands, \textsl{(2)} Ca sphere, \textsl{(3)} internal $\pi$ bands (the one cutting the Ca sphere, as shown in
Fig.~\ref{fig:CaC6bands}). We obtained:

\vspace{0.4cm}
\begin{center}
\begin{minipage}{\textwidth}
\begin{center}
\begin{minipage}{0.36\textwidth}
\begin{center}
$\lambda: $

\vspace{0.2cm}
$\left[\begin{array}{ccc} 0.286 & 0.173 & 0.223 \\  
                                   0.518 & 0.315 & 0.425 \\
                                   0.382 & 0.245 & 0.303 \end{array}\right]$
\end{center}
\end{minipage}
\begin{minipage}{0.36\textwidth}
\begin{center}
$V^{ph}\left[eV\right]:$

\vspace{0.2cm}
$\left[\begin{array}{ccc} 0.363 & 0.651 & 0.492 \\
                            0.651 & 1.177 & 0.932 \\
                            0.492 & 0.932 & 0.678 \end{array}\right]$
\end{center}
\end{minipage}
\begin{minipage}{0.24\textwidth}
\begin{center}
${\rm DOS}\left[\frac{\rm states}{eV\cdot{\rm cell}}\right] :$

\vspace{0.2cm}
$\left[\begin{array}{c} 0.79  \\ 0.27  \\ 0.45 \end{array}\right]$
\end{center}
\end{minipage}
\end{center}
\end{minipage}
\end{center}

\vspace{0.4cm}

For MgB$_2$ we have \cite{MgB2-Review}: 

\vspace{0.4cm}
\begin{center}
\begin{minipage}{\textwidth}
\begin{center}
\begin{minipage}{0.36\textwidth}
\begin{center}
$\lambda:$

\vspace{0.2cm}
$\left[\begin{array}{cc} 1.017 & 0.213  \\
                                  0.156 & 0.448  \end{array}\right]$
\end{center}
\end{minipage}
\begin{minipage}{0.36\textwidth}
\begin{center}
$V^{ph}\left[eV\right]:$

\vspace{0.2cm}
$\left[\begin{array}{cc} 3.390 & 0.520 \\
                                          0.520 & 1.093 \end{array}\right]$
\end{center}
\end{minipage}
\begin{minipage}{0.24\textwidth}
\begin{center}
${\rm DOS}\left[\frac{\rm states}{eV\cdot{\rm cell}}\right] :$ 

\vspace{0.2cm}
$\left[\begin{array}{c} 0.30 \\ 0.41 \end{array}\right]$
\end{center}
\end{minipage}
\end{center}
\end{minipage}
\end{center}

\vspace{0.4cm}
\noindent where the first entry in each row or column corresponds to the two $\sigma$ bands and the second to  the $\pi$ \cite{Golubov}.   

These data show clearly that both systems are quite anisotropic: In MgB$_2$, the much stronger coupling in
the $\sigma$ rather than in the  $\pi$ Fermi surfaces and the small interband coupling, makes the
calculated anisotropic \tc\ double with respect to the isotropic one \cite{noimgb2,MgB2-Review}.
Within the Suhl-Matthias-Walker model\cite{SMW} this can be roughly explained by a large $\lambda_{max}$=1.070 (defined as the maximum eigenvalue of the $\lambda_{ij}$ matrix - very similar to $\lambda_{11}$) with respect to the average, isotropic coupling $\bar\lambda$=0.860. 

In CaC$_6$, instead, both Ca and C bands contribute strongly to the global coupling. 
Consider the Ca spherical FS, that in CaC$_6$ has the largest superconducting gap. As shown by the
$\lambda_{ij}$ matrix, the contribution coming from the interband scattering is larger than the intraband
one ($\lambda_{21}$=0.518 and $\lambda_{23}$=0.425, against the intraband $\lambda_{22}$=0.315): Although the Ca intraband average coupling (as represented by the $V_{ij}^{ph}$  matrix) is the largest one, the large phase space (DOS) for interband scattering, with a reasonably large interband scattering potential, makes the difference. 
We calculated $\lambda_{max}$=0.912 and $\bar\lambda$=0.855, both much larger than any of the intraband couplings. 
These arguments are supported by the SCDFT result: Using the isotropic approximation we obtain a  critical temperature of 8.1 K, against \tc= 9.4 K obtained within the fully anisotropic approach. Hence, unlike in MgB$_2$, we have only a  15\% reduction on \tc\ in comparison with the anisotropic case.

\section{H under pressure}\label{results_H}

The possibility of a superconducting dense molecular phase of hydrogen 
represents a long-standing problem, recently investigated by means of
first-principles methods within the SCDFT\cite{noiH}.
The low temperature and high pressure ($>$ 400 GPa) phase of
molecular hydrogen is predicted to be a base-centered orthorhombic metallic molecular solid
 (known as $Cmca$ phase)\cite{needs} with two molecules per unit cell
 located on different layers.
 The electronic band structure arises from the bonding and anti-bonding
 combination of the H$_2$ molecular orbitals. 
 At high pressure, the band overlap between the valence and
 conduction bands produces a  complex Fermi surface with 
 disconnected sheets of different orbital nature\cite{noiH}.
The molecular nature of the bands provides a strong
electron-phonon coupling.
The superconducting properties of this system are extremely interesting: The
calculated \tc\ results to be around 240 K at $\simeq$ 450 GPa,
with three different  gaps on the three different sheets of 
the Fermi surface.
The presence of three gaps represents one of the main
peculiarities of molecular hydrogen, and its origin is mainly due to the (band) anisotropy of the electron-phonon coupling \cite{noiH}.
However, as we will see, the 
anisotropy of the Coulomb interaction is fundamental to achieve 
a proper prediction of the gaps at T=0 K.

In order to simplify the analysis, we will use a multigap BCS model (with parameters calculated {\it ab-initio}) to discuss the effect of multiband
anisotropy, with particular attention to the Coulomb interactions.
Thus, we reduce the fully anisotropic SCDFT gap equation to a BCS
multiband equation, including phonon renormalization effects [which, in the SCDFT gap Eq. (\ref{eq:gap}) are included in the term ${\cal Z}_\nk$].
In order to do this, we need the band-resolved interaction matrices, calculated
averaging the {\bf k}-resolved e-ph and e-e matrix elements over (physically) 
different sheets of the Fermi surface.   
 
\begin{figure}
\includegraphics[width=0.5\textwidth]{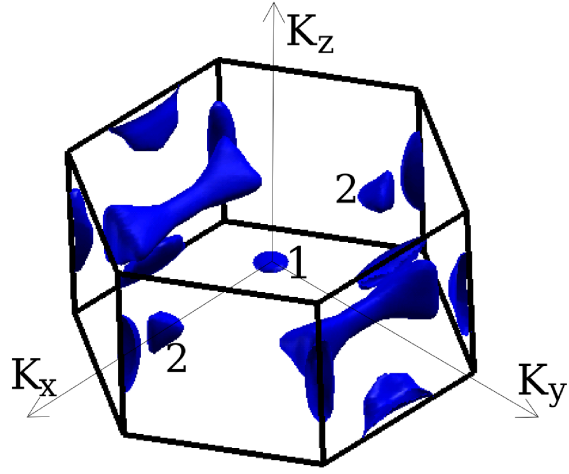}
\caption{Fermi surface of molecular hydrogen at 414 GPa. The different sheets
used to construct the interaction matrices are labeled as  (1): the disk region;
(2): the "prism-like" regions; the remaining regions (not labelled).}
\label{fermi}
\end{figure}
In Fig.\ref{fermi} we show the calculated Fermi surface at 414 GPa. The three
gaps separate into three main regions: the disk at the Gamma point (labeled as 1), the "prism-like" regions (2) and the remaining.
In order to construct a 3-band model of the system, we introduce a 3$\times$ 3
 $\lambda$-matrix of the partial e-ph coupling $\lambda_{ij}$ where
the indices $i,j=1,2,3$ span over the above mentioned regions.
We also introduce  the 3$\times$3 matrix for the $V^{ph}$, as defined in Sec. \ref{effph}.
In analogy with the e-ph interaction, we introduce also the  3$\times$ 3
Coulomb interaction matrix $V_{ij}^{el}$ and the $\mu_{ij}$ matrix defined by the relation:
$\mu_{ij}=N_{j}V_{ij}^{el}$. $V_{ij}^{el}$ represents the band resolved
Coulomb matrix defined in analogy with the definition of $\lambda_{ij}$, but containing the Coulomb matrix elements.
In the following, we report the values of these  matrices:

\vspace{0.4cm}

\begin{center}
\begin{tabular}{cc}

$V^{ph}\left[Ry\right]:$  &  $V^{el}\left[Ry\right]:$ \\

$\left[\begin{array}{ccc}          2.16  & 3.00 &  1.59 \\  
                                   3.00 & 0.45 & 0.91 \\
                                   1.59 & 0.91 & 0.78 
\end{array}\right]$  &

$\left[\begin{array}{ccc}   0.71 & 0.08 & 0.11 \\
                           0.08 & 0.38 & 0.20 \\
                           0.11 & 0.20 & 0.20 
\end{array}\right]$ \\

   & \\

$\lambda:$          &     $\mu:$     \\

$\left[\begin{array}{ccc}          0.14  & 0.37 &  1.45 \\  
                                   0.20 & 0.06 & 0.82 \\
                                   0.11 & 0.11 & 0.71 
\end{array}\right]$ &

$\left[\begin{array}{ccc}   0.046 & 0.010 & 0.100 \\
                           0.006 & 0.046 & 0.180 \\
                           0.007 & 0.025 & 0.182
\end{array}\right]$ \\

\end{tabular}
\end{center}

We note that the intraband scattering ($V_{ii}^{el}$) is dominating the Coulomb interaction
 and that it is particularly strong for the  states of  
region (1). These states are bonding combinations of anti-bonding molecular
states,  are responsible for the inter-layer bonding and are very
localized in the inter-layer region, with rather poor screening properties.
An opposite behavior characterizes the $\mu$ matrix, which represents the
effective e-e interaction in the superconducting phase: in fact, despite the
large value of $V_{11}^{el}$ we find a very small value of $\mu_{11}$, due 
to the low density of states in region (1). 
Interband terms, in particular those involving the bands less coupled with
phonons, dominate thanks to their high density of states.
Although  inspection of the $\mu$ matrix suggests a minor role 
of the e-e interaction anisotropy  in the  \tc\ value
(we find the same T$_c$ in the isotropic and anisotropic
case), we demonstrate that inclusion of band anisotropy in the repulsive term
has a non trivial effect on the T=0~K gap.
The phonon renormalized BCS equation (RBCS) is obtained from the SCDFT fully anisotropic
equation averaging the $\bf{k}$-anisotropy (still retaining the band anisotropy)
and performing the $T\rightarrow 0$, 
$\xi, \xi' \rightarrow 0$ limits. 
The results of the solution of the RBCS equation are reported in Table I.
\begin{table}[h]
\label{resu}
\begin{tabular}{|c|c|c|}
\hline
              & $ \Delta_1/\Delta_3 $ &  $\Delta_2/\Delta_3$ \\
\hline
Full-RBCS          &	1.61              &  1.13               \\ 
RBCS (e-ph)  &     1.37              &  1.09              \\
\hline
\end{tabular}
\caption{Results of the RBCS equation including  both the e-ph ($\lambda_{ij}$) 
and e-e ($\mu_{ij}$) matrices (Full-RBCS), and without the $\mu_{ij}$ matrix [RBCS (e-ph)].} 
\end{table}  

As we can see from Table I the presence of three gaps is predicted
even within the RBCS model.
Comparing the results of the Full-RBCS and RBCS (e-ph),  the role
of both interaction on the prediction of T=0 gaps is evident. Even if the main source of
band anisotropy is given by the e-ph interaction,  inclusion of the
proper band-resolved Coulomb interaction increases the anisotropy of about 18\%. 

This demonstrates that the band anisotropy of the repulsive 
e-e interaction, essentially related to the different nature 
of the bands at the Fermi level, increases the multigap superconducting properties in 
dense molecular hydrogen. 

\section{Summary}\label{concl}

The SCDFT approach allows  a 
fully anisotropic description of the superconducting phase and  the inclusion 
of Coulomb repulsion  effects on an {\it ab-initio} basis.  
After a brief review of the SCDFT method, in this paper we performed a
detailed analysis of the Coulomb and electron-phonon matrix elements 
in intercalated graphite \cac\ and hydrogen under high pressure.
In \cac\ we studied different approximations for 
the Coulomb interaction. We find that, 
due to the presence of strongly localized states, 
the use of free electron-like Coulomb matrix elements gives 
a strong underestimation of the Coulomb repulsion and a corresponding 
overestimation of the critical temperature \tc. 
While local field effects 
are important to describe the 
screening properties in the graphene layers,  
inclusion of  an isotropic Coulomb interaction -
averaged on isoenergy surfaces - does not strongly affect the value of \tc. Concerning the electron-phonon interaction, instead, the isotropic approximation yields a \tc\ reduction of 15\%.

Finally, we calculate the matrix elements of both interactions for  H under high pressure. By making use of  a multiband BCS model using SCDFT-calculated parameters, we demonstrate that the usual isotropic approximation of the
repulsive e-e interaction is completely unjustified in the case of
superconducting molecular hydrogen, and that the Coulomb anisotropy it is fundamental for the prediction of the three superconducting gaps at T=0 K.

\section{Acknowledgments}
This work makes use of results produced by the Cybersar Project 
managed by the Consorzio COSMOLAB,   
co-funded by the Italian Ministry of University and Research 
(MUR) within the Programma Operativo Nazionale 2000-2006 
"Ricerca Scientifica, Sviluppo Tecnologico, Alta Formazione" 
per le Regioni Italiane 
dellÕObiettivo 1  Ð Asse II, Misura II.2 ÒSocietˆ dellÕInformazioneÓ.
Work partially supported by the Italian Ministry of
Education, through PRIN 200602174 project, by INFM-CNR
through a supercomputing grant at Cineca (Bologna, Italy),
by the Deutsche Forschungsgemeinschaft and by
NANOQUANTA Network of Excellence.


\end{document}